\documentclass[wssquare]{ws-procs975x65} 

\def\be{\begin{equation}}
\def\ee{\end{equation}}
\newcommand{\bea}{\begin{eqnarray}}
\newcommand{\eea}{\end{eqnarray}}
\def\l{\lambda}
\def\L{\Lambda}
\def\la{\lambda}
\def\m{\mu}
\def\n{\nu}

\def\pa{\partial}
\def\t{\tau}

\def\b{\beta}
\def\k{\kappa}
\def\e{\epsilon}
\def\xBZ{x_\mathrm{BZ}}
\newcommand{\nn}{\nonumber}
\def\order#1{$\mathcal{O}\left(#1\right)$}
\def\morder#1{\mathcal{O}\left(#1\right)}

\def\href#1#2{#2}

\begin{document}
\title{Holography and the conformal window in the Veneziano limit}

\author{M. J\"arvinen$^*$}

\address{Laboratoire de Physique Th\'eorique,
\'Ecole Normale Sup\'erieure \&\\ Institut de Physique Th\'eorique Philippe Meyer,\\ 24 rue Lhomond, 75231 Paris Cedex 05, France\\
$^*$E-mail: jarvinen@lpt.ens.fr}

\begin{abstract}
We discuss holographic QCD in the Veneziano limit
(the V-QCD models), 
concentrating on phenomena near the ``conformal'' phase
transition taking place at a critical value of the ratio $x\equiv N_f/N_c$.  In particular, we review the
results for the S-parameter, the technidilaton, and the masses of
the mesons. 
\end{abstract}


\bodymatter

\section{Introduction and motivation}

For QCD in the Veneziano limit~\cite{Veneziano:1979ec},
\be \label{Vlimit}
 N_c \to \infty\ , \quad N_f \to \infty \ , \quad x \equiv \frac{N_f}{N_c}\ \ \mathrm{fixed} \ , \quad g^2 N_c \ \ \mathrm{fixed}\ ,
\ee
the number of quarks ($\sim N_f N_c$) and the number of gluons ($\sim N_c^2$) are comparable. Consequently, the flavor degrees of freedom are fully dynamical and there is a nontrivial interplay between the glue and flavor sectors of the theory. This backreaction of the quarks to the dynamics of the gluons would be absent in the 't Hooft limit, which is the same as~\eqref{Vlimit} except that $N_f$ is kept fixed instead of $x$ such that $N_f \ll N_c$. Therefore studying QCD in the Veneziano limit is more complicated than in the 't Hooft limit, but it is well motivated because there are interesting phenomena which cannot be captured if $N_f \ll N_c$. In particular, the phase diagram of QCD is nontrivial as a function of the variable $x$ which is continuous in the Veneziano limit. 

\begin{figure}[!tb]
\begin{center}
\includegraphics[width=0.5\textwidth]{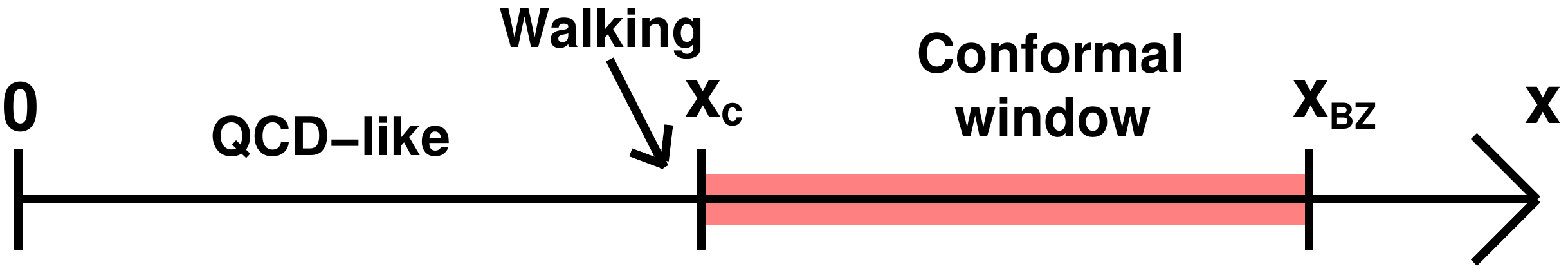}
\end{center}
\caption{The conjectured phase diagram of QCD in the Veneziano limit (at zero temperature and quark mass) as a function of $x=N_f/N_c$.}
\label{fig:xphases}\end{figure}

The ``standard'' expectation for the phase diagram as a function of $x$ (at zero temperature and quark mass) is shown in Fig.~\ref{fig:xphases}. Within the interval $0<x< 11/2 \equiv \xBZ$, where the theory is asymptotically free, the following regimes can be identified:
\begin{itemize}
 \item \emph{The QCD regime} with $0<x<x_c$, where the infrared (IR) dynamics is similar to ordinary QCD (i.e., a theory with $N_c=3$ and a few light quarks). That is, the theory is confining and breaks chiral symmetry.
 \item \emph{The walking regime} with $0<x<x_c$ and $x_c-x \ll 1$, where the coupling constant of the theory varies very slowly, i.e., ``walks'', over a large range of energies.
 \item \emph{The conformal window} with $x_c<x<x_\mathrm{BZ}$, where the theory runs to an IR fixed point (IRFP). 
\end{itemize}

The existence of the conformal window is guaranteed in the Banks-Zaks (BZ) limit $x\to \xBZ$, because the value of the coupling is parametrically small for the entire renormalization group (RG) flow and perturbation theory is trustable. It is also credible that going to large $N_c$ and $N_f$ at small $x \lesssim 1$ does not change the dynamics much, and it is therefore similar to that of ordinary QCD. But the nature of the ``conformal transition'' at $x=x_c$, and the behavior of the theories near the transition, is much more difficult to study and remains essentially an open question.

The phase diagram of Fig.~\ref{fig:xphases} with a walking regime is obtained in the Dyson-Schwinger approach~\cite{Appelquist:1988yc}. The transition is then of the Berezinsky-Kosterlitz-Thouless (BKT) type, associated with the so-called Miransky scaling law. The existence of the walking regime is important, since theories in this region may have properties, which are desirable for technicolor candidates. 
However, it has also been suggested that the transition is discontinuous~\cite{Sannino:2012wy}.
There is an ongoing effort to study the theories near the transition by using lattice simulations~\cite{Neil:2012cb,Giedt:2012it,Kuti:2014epa,Lombardo:2014mda}, but as it turns out, obtaining reliable results is much more difficult than for ordinary QCD.

Here we discuss the conformal transition and related by using holography and working in the Veneziano limit. We concentrate on a specific bottom-up approach, which tries to follow  principles from string theory as closely as possible. That is, the model is derived by using five dimensional noncritical string theory with a certain brane configuration. However, as it turns out, some things do not work -- in particular the ultra violet (UV) physics is not that of QCD. This can be fixed by modifying the various potentials of the model by hand, therefore switching from a top-down to a bottom-up approach. As the potentials are thereafter not those derived from string theory, the model should be understood as an effective framework which could in principle be used to model more general field theories than just QCD, depending on the choice of potentials. It makes sense to fit the remaining degrees of freedom in the potentials to QCD data, picking a specific model which has physics similar to QCD. When working in the Veneziano limit, this kind of approach leads to the V-QCD models~\cite{Jarvinen:2011qe} (where ``V'' stands for Veneziano). The V-QCD models also have the phase diagram of Fig.~\ref{fig:xphases}, including a BKT transition and walking.

In more detail, V-QCD is based on two building blocks, the improved holographic QCD (IHQCD)~\cite{Gursoy:2007cb,Gursoy:2007er} for the gluon sector, and a method for adding flavor by inserting a pair of space filling $D4-\overline{D4}$ branes~\cite{Bigazzi:2005md,Casero:2007ae}. 
The glue and flavor sectors are fully backreacted in the Veneziano limit. The structure of V-QCD at finite temperature and chemical potential~\cite{Alho:2012mh,Alho:2013hsa,Iatrakis:2014txa,Alho:2015zua} as well as the two-point correlators and bound state masses~\cite{Arean:2012mq,Arean:2013tja,Jarvinen:2015ofa} have also been analyzed.

In this talk we will first discuss IHQCD and V-QCD in more detail. Then we will give an overview of the results for V-QCD. We will concentrate on the physics related to the conformal transition and the conformal window -- for a shorter and more general review on V-QCD see~\cite{Jarvinen:2015wia}.

\section{Building blocks of V-QCD}

\subsection{Improved holographic QCD}

Improved holographic QCD (IHQCD)~\cite{Gursoy:2007cb,Gursoy:2007er} is a model for the Yang-Mills theory, which is loosely based on five dimensional noncritical string theory. The most relevant terms of the IHQCD action, which appear at leading order in the large $N_c$ ('t Hooft) limit, are
\be \label{LIHQCD}
S_\mathrm{IHQCD}= M^3 N_c^2 \int d^5x  \sqrt{-\det g}\left(R-\frac{4}{3}{
(\partial\phi)^2}+V_g(\phi)\right)\,,
\ee
where the first term defines the five dimensional Einstein gravity and $\phi$ is the dilaton. 

The dilaton is linked to the 't Hooft coupling $g_s^2 N_c$. That is, we denote $\l = e^{\phi}$ and identify the vacuum solution $\l(r)$ (a function of the bulk coordinate $r$), which extremizes of the action~\eqref{LIHQCD}, as the (running) 't Hooft coupling. The Ansatz for the metric is
\be
ds^2=e^{2 A(r)} (dx_{1,3}^2+dr^2)\,.
\label{metric1}
\ee
Here $A$ is identified as the logarithm of the energy scale in field theory and can therefore be used to define the holographic RG flow. In our
conventions the UV boundary lies at $r=0$, and the bulk coordinate runs from zero to infinity. The
metric will be close to the AdS near the UV ($A \sim -\log(r/\ell)$, where $\ell$ is the UV AdS radius). Therefore, $r$ corresponds roughly to the inverse
of the energy scale of the dual field theory.

As we already pointed out, the model deviates from the ``top-down'' derivation as the dilaton potential $V_g$ is fixed by hand rather than derived from string theory. Its asymptotic behavior is determined by matching with general properties Yang-Mills theory as follows.

In the UV (i.e, as $\l \to 0$), $V_g$ must asymptote to a constant, in order for the geometry to be asymptotically AdS. Moreover, if $V_g$ has an analytic expansion at $\l=0$, the coupling $\l$ turns out to have logarithmic dependence on $r$ in the UV, reminiscent of the RG flow of the coupling in Yang-Mills. Indeed, the holographic beta function defined by the above dictionary
\be \label{hbeta}
 \beta = \frac{\l'(r)}{A'(r)}
\ee
admits a Taylor expansion at $\l=0$ (with $\l=\l(r)$ evaluated on the vacuum solution). In order to fix $V_g$ in the UV, the coefficients in its Taylor expansion are then chosen such that the holographic beta function~\eqref{hbeta} agrees with the perturbative beta function of Yang-Mills up to two-loop order. It may sound weird that a holographic model is matched with perturbative QCD, since holography typically cannot be used easily at weak coupling. This is however the best we can do to include the UV physics of QCD, and should provide as correct boundary conditions as possible for the more interesting IR dynamics in the model.

In the IR (i.e, at large $\l$), we require the potential to diverge as 
\be
 V_g\sim \l^{4 / 3}\sqrt{\log\la}\ , \qquad  (\l\to \infty)\ .
\ee
This ensures confinement,
a mass gap, discrete spectrum and asymptotically linear glueball trajectories (for radial excitations) \cite{Gursoy:2007cb,Gursoy:2007er,Gursoy:2009jd}. Interestingly, this asymptotic behavior matches with the behavior predicted by string theory $V_g \sim \l^{4/3}$ up to the logarithmic correction, which is important to ensure that the glueballs have exactly linear trajectories.

\begin{figure}[tb]
\centering
\includegraphics[width=0.5\textwidth]{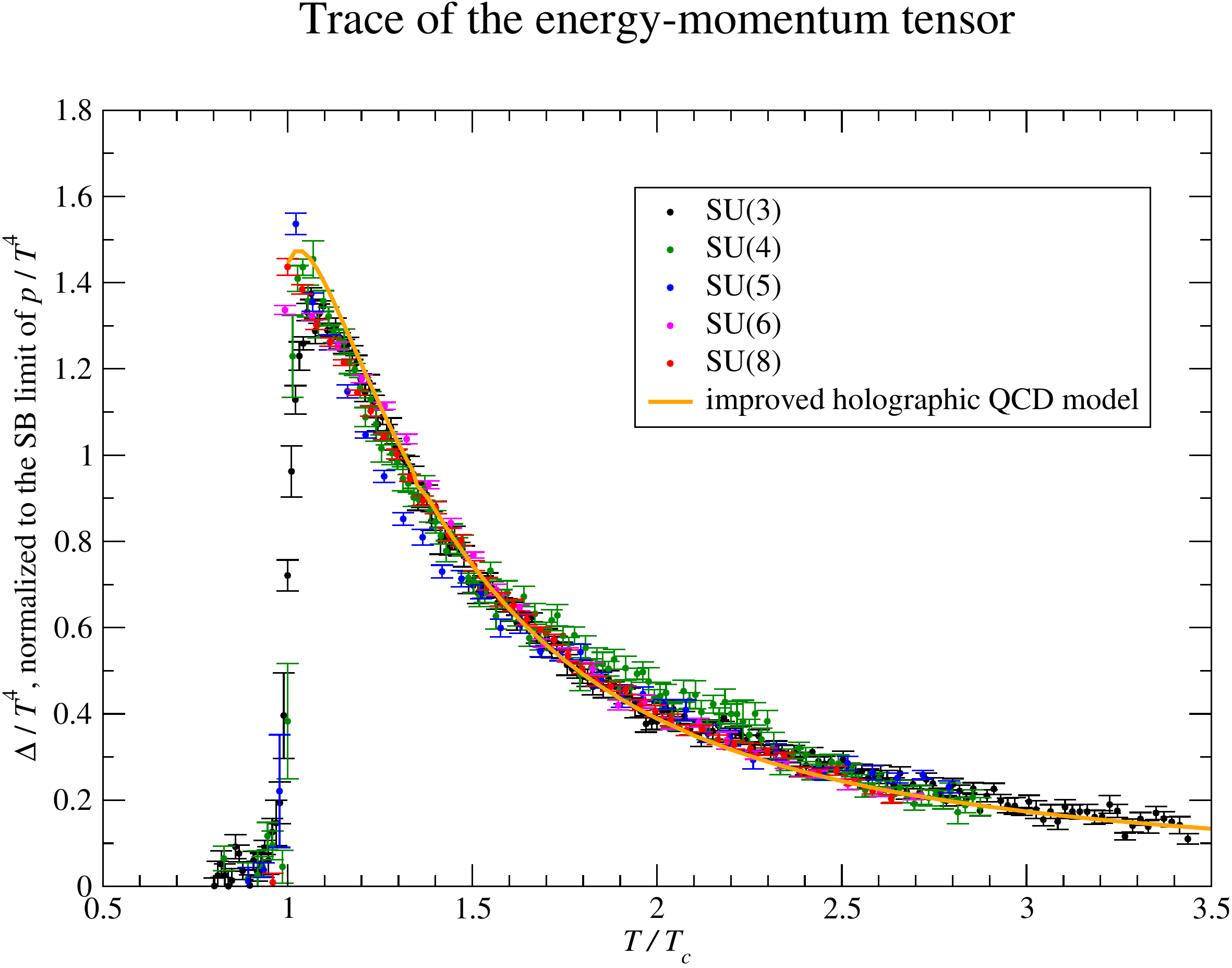}
\caption{The interaction measure $(p-3\e)/T^4$ 
in IHQCD compared to Yang-Mills lattice data at various number of colors~\protect\cite{Panero:2009tv}.}
\label{fig:Ttrace}
\end{figure}

The behavior of $V_g$ at intermediate couplings, $\l = \morder{1}$, must be fitted to Yang-Mills data. Most precise constraints are obtained by studying the model at finite temperature~\cite{Gursoy:2008bu,Gursoy:2008za}. An example 
of a two parameter fit~\cite{Gursoy:2009jd} to lattice data for the interaction measure in $SU(3)$ Yang-Mills is given in Fig.~\ref{fig:Ttrace}, compared to lattice data at higher $N_c$ which was computed afterwards~\cite{Panero:2009tv}.

\subsection{Adding flavor}

A framework for adding flavor in holographic models has been introduced in~\cite{Bigazzi:2005md,Casero:2007ae}. A pair of $D4-\overline{D4}$ branes is added, which fill the whole space in a five dimensional holographic model. The string stretching between the branes gives rise to a tachyon field, which is dual to the $\bar q q$ operator on the field theory side, and therefore controls chiral symmetry breaking. The gauge fields on the branes are dual to the left and right handed currents $\bar q (1\pm\gamma_5)q$ in QCD. 

The dynamics of the flavor is governed by a tachyon-Dirac-Born-Infeld (TDBI) action, and taking a Sen-like exponential potential, $V_f(\t) =\exp(-\t^2)$, for the tachyon. This framework has been tested in the probe approximation ($x \to 0$ limit) for IHQCD~\cite{Gursoy:2007er} and in more detail~\cite{Iatrakis:2010jb} for the Kuperstein-Sonnenschein background~\cite{Kuperstein:2004yf} with encouraging results. Confining background was seen to trigger chiral symmetry breaking, the pion modes were identified and their mass were shown to obey the Gell-Mann-Oakes-Renner relation, and a good description of the masses of the light mesons in QCD was obtained.

\section{V-QCD}

The V-QCD models arise as the fusion of IHQCD with the framework for adding flavor in the Veneziano limit where flavor fully backreacts to the glue~\cite{Jarvinen:2011qe}.
The main degrees of freedom are therefore the two scalars: the ``dilaton'' $\l=e^\phi$ and the tachyon $\t$. The dictionary is as stated above, in particular (the background value of) $\l$ is the 't Hooft coupling, and the tachyon is dual to the $\bar q q$ operator.

The terms in the V-QCD action which determine the vacuum solution (for which the gauge fields on the branes vanish) are
\begin{align} \label{LVQCD}
S_\mathrm{VQCD}&= M^3 N_c^2 \int d^5x \sqrt{-\det g}\left(\!R-\frac{4}{3}\frac{\left(\pa \l\right)^2}{\l^2}+V_g(\l)\!\right) \nn\\
 & \quad - M^3 N_f N_c \int d^5x\ V_f(\l,\t) \sqrt{-\det\left(g_{\m\n} + \k(\l)\pa_\mu\t\pa_\n\t\right)} \ .
\end{align}
The first term is the IHQCD action~\eqref{LIHQCD} and describes the dynamics of the glue. The second term is a generalized TDBI action, which describes the dynamics of the flavor. From the explicit dependence on $N_c$ and $N_f$ we see that both terms are of the same order in the Veneziano limit and there is indeed full backreaction between the glue and flavor sectors. 

Notice that the form of the TDBI action is only known in the probe limit $x \to 0$. Therefore we have introduced a more general DBI type action in~\eqref{LVQCD}, which is consistent with the probe limit action. For the tachyon potential we take
\be 
 V_f(\l,\t) = V_{f0}(\l)e^{-a(\l)\t^2}
\ee
where the dependence on the tachyon is the same as in the probe limit approach~\cite{Casero:2007ae}. Notice that the functions $V_{f0}(\l)$, $\k(\l)$, and $a(\l)$ would be power laws in the probe limit, but we have taken them to be (at this point) arbitrary functions of $\l$. This generalizes the approach of IHQCD, where the dilaton potential $V_g$ was taken to be a more complicated function of $\l$ than the power law predicted by string theory analysis, to the case with backreacted flavors.

We need to specify the functions $V_g(\l)$, $V_{f0}(\l)$, $\k(\l)$, and $a(\l)$ in order to pin down the model. It turns out that the dilaton potential $V_g(\l)$ must obey the same asymptotic constraints which were discussed above for IHQCD. For the other functions similar constraints are found, both in the UV and in the IR.

In the UV ($\l \to 0$) the potentials must approach constant values in order for the dimension of the $\bar q q$ operator to be correctly reproduced, among other things. 
In analogy to the approach taken in IHQCD, we take the potentials to be analytic at $\l=0$ and match the coefficients of their Taylor expansions with the perturbative QCD. As it turns out, the coefficients of $V_{f0}(\l)$ can be mapped to the ($x$ dependent) coefficients of the QCD beta function, and the coefficients of the ratio $a(\l)/\k(\l)$ are mapped to the coefficients of the anomalous dimension of the quark mass $\gamma = -d\log m_q/d \log \m$. 
The result from this mapping is that the tachyon has the following asymptotics in the UV ($r \to 0$):
\be \label{tachUV}
 \t(r) \simeq  m_q\ r\ (-\log r \L)^{-\gamma_0/\b_0}  + \langle \bar qq\rangle\ r^3\  (-\log r \L)^{\gamma_0/\b_0}
\ee
where $\beta_0$ and $\gamma_0$ are the leading coefficients of the beta and gamma functions, respectively, and the scale of the holographic RG flow $\L$ corresponds to $\Lambda_\mathrm{QCD}$ on the field theory side. The implemented RG flow (the logarithmic terms) is the same as the leading perturbative RG flow of the quark mass and the condensate in QCD.
Therefore the UV matching is done at slightly higher level than in the traditional bottom-up models. As we stressed above for IHQCD, it should be understood that this matching is done in order to guarantee correct boundary conditions for the IR dynamics, which can be more reliably modeled by holography.

In the IR ($\l \to \infty$), the functions are constrained by the asymptotics of the spectra at large excitation numbers, and by requiring that the IR geometry has a ``good'' kind of IR singularity so that  all IR boundary conditions are uniquely fixed. These conditions lead to the following asymptotics for $a$ and $\kappa$~\cite{Arean:2013tja}:
\be \label{akasympt}
\begin{split}
 a(\l) &\sim \mathrm{const.}\\
 \k(\l) &\sim \l^{-4/3}\sqrt{\log \l} 
\end{split}\  \qquad (\l \to \infty) \ .
\ee
Interestingly, these asymptotics agree with the string theory prediction, up to the logarithmic correction in $\k(\l)$, in analogy to what was found for $V_g(\l)$ in IHQCD. The function $V_{f0}(\l)$ is less constrained, but must diverge asymptotically in the IR.

Finally, the potentials at intermediate values of $\l$ could be fitted to lattice and experimental QCD data (such as the equation of state and meson masses) in order to build a concrete model for QCD. An overall fit is the topic of ongoing work. The results presented below were computed with potentials obeying the asymptotics found above but were not fitted to data.

\section{Properties of V-QCD}

\subsection{Background solutions}

In order to study the phase structure of V-QCD, we take an Ansatz where the fields $\l(r)$ and $\t(r)$ depend on the bulk coordinate $r$, and the metric has the form~\eqref{metric1}. The Ansatz is then inserted to the equations of motion derived from the action~\eqref{LVQCD}, which are solved numerically. The free energy of each solution is computed in order to identify the dominant solution.

This procedure results in a zero temperature phase
diagram that is essentially universal and the same as that of Fig.~\ref{fig:xphases}. By universality, we mean that the phase diagram is largely independent on the exact choice of the potentials in the V-QCD action, and qualitatively similar to all potentials which smoothly interpolate between the UV and IR asymptotics discussed in the previous section. In addition, many results near the conformal transition hold for a little more general class of models than just V-QCD~\cite{Jarvinen:2015ofa} as we will explain below.

As a function of $x$ and at zero quark mass
there are two phases (restricting to the range $0<x<\xBZ$) separated by a phase transition at some $x=x_c\simeq 4$:
\begin{itemize}
 \item In the region
 $0<x<x_c$, the theory breaks chiral symmetry, and IR dynamics is similar to that of ordinary QCD. The background solution for V-QCD
 has nontrivial $\l(r)$, $A(r)$ and $\tau(r)$ in this region, ending at the ``good'' kind of singularity in the IR.
 \item In the conformal window, i.e., when $x_c<x<{\xBZ}$, the theory flows to a nontrivial IR fixed point and there
 is no chiral symmetry breaking.
  The background solution has zero tachyon $\tau(r)=0$ (so that indeed the quark mass and condensate in~\eqref{tachUV} vanish) and nontrivial $\l(r)$ and $A(r)$, giving rise to a
 geometry which is asymptotically AdS both in the UV and in the IR~\cite{Jarvinen:2009fe}.
\end{itemize}
As $x \to x_c$ from below, ``walking'' RG flow of the coupling constant is found: the background flows very close to the IRFP (i.e., the geometry is close to AdS) but eventually the nonzero value of the tachyon drives the system away from the fixed point very deep in the IR.

\subsection{BF bound and the conformal transition}

The existence of the perturbative fixed point in the Banks-Zaks region (as $x \to \xBZ$ from below) is a consequence of matching the potentials to the perturbative RG flow of QCD in the UV. The conformal transition and the walking regime, on the contrary, are generated by the dynamics of the model in a nontrivial way. Let us discuss the holographic mechanism leading to this structure.

The conformal transition is related to an instability of the tachyon field in the conformal window~\cite{Kaplan:2009kr}, which arises when the five dimensional bulk mass of the tachyon, evaluated at the IR fixed point, violates the so-called Breitenlohner-Freedman (BF) bound. That is, an infinitesimal tachyon field near an AdS fixed point obeys the flow
\be
 \tau(r) \simeq C_1 r^{\Delta_*} + C_2 r^{4-\Delta_*} \ .
\ee
Here the exponents are related to the squared bulk mass of the tachyon $m_\tau$ and the AdS radius $\ell_*$ as
\be
 - m_\tau^2 \ell_*^2 = \Delta_*(4-\Delta_*) \le 4
\ee
where the inequality, the BF bound, holds for real $\Delta_*$. When the bound is violated, $\Delta_*$ becomes complex, leading to an oscillating tachyon solution which signals an instability towards forming a tachyon condensate in the bulk, and therefore a chiral condensate in field theory through~\eqref{tachUV}. Therefore the point where the bound is saturated marks the location of the conformal transition $x=x_c$. At the transition, the anomalous dimension of the quark mass $\gamma_*=\Delta_*-1 =1$ which agrees with the expectation from Dyson-Schwinger analysis.

\begin{figure}[tb]
\centering
\includegraphics[width=0.5\textwidth]{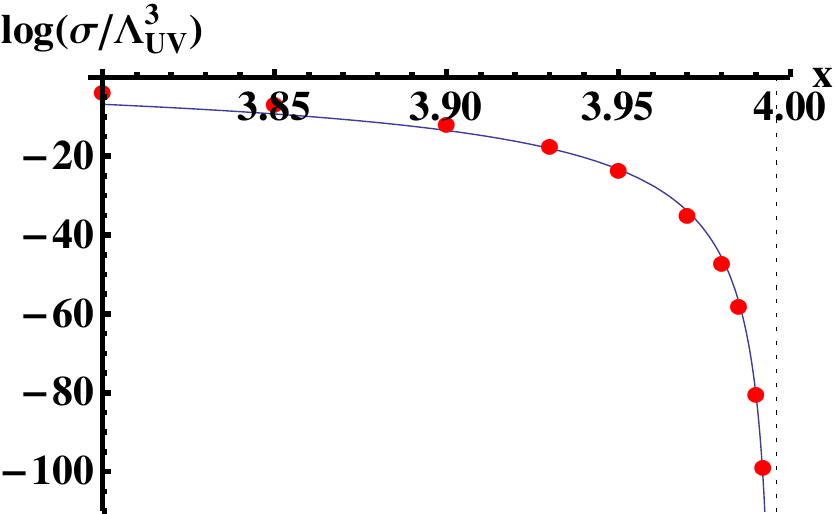}
\caption{The dependence of the chiral condensate on $x$ near the conformal transition. The blue dots are data from V-QCD, and the red curve is a fit by the Miransky scaling law~\eqref{condscaling}.}
\label{fig:condscaling}
\end{figure}

Transition triggered by the violation of the bound quite in general leads to walking and a BKT/Miransky phase transition of infinite order. This transition involves a characteristic exponential scaling law, called the Miransky scaling. The chiral condensate, for example, vanishes as
\be \label{condscaling}
 \langle \bar qq\rangle \sim \exp\left(-\frac{2 K}{\sqrt{x_c-x}}\right)\ ,
\ee
as $x \to x_c$ from below, see Fig.~\ref{fig:condscaling}. The coefficient $K$ is given by
\be
 K =\frac{\pi}{\sqrt{\frac{d}{dx}\left[m_\tau^2 \ell_*^2\right]_{x=x_c}}}\ .
\ee

The presence of the Miransky scaling also means that the theory involves two energy scales and a large scale hierarchy in the walking regime. We denote by $\L=\L_\mathrm{UV}$ the UV scale which already appeared in~\eqref{tachUV} and maps to $\L_\mathrm{QCD}$ on the field theory side. The other scale is denoted by $\L_\mathrm{IR}$. It gives the scale of soft nonperturbative physics and for example the rho mass in the walking regime. The relation between the scales is given by
\be
 \frac{\L_\mathrm{UV}}{\L_\mathrm{IR}} \sim \exp\left(\frac{K}{\sqrt{x_c-x}}\right)\ , \qquad (x_c-x \ll 1)\ .
\ee

Notice that these arguments do not use directly the action of V-QCD. Indeed, as we already commented above, the physics near the transition is largely independent of the details of the holographic model. In particular, conformal transition (of the BKT type) has been found and analyzed earlier in other holographic models and the behavior near the transition is similar to that of V-QCD in these models~\cite{Kutasov:2011fr,Kutasov:2012uq,Goykhman:2012az,Alvares:2012kr,Evans:2014nfa}.

\begin{figure}[!tb]
\centering
\includegraphics[width=0.45\textwidth]{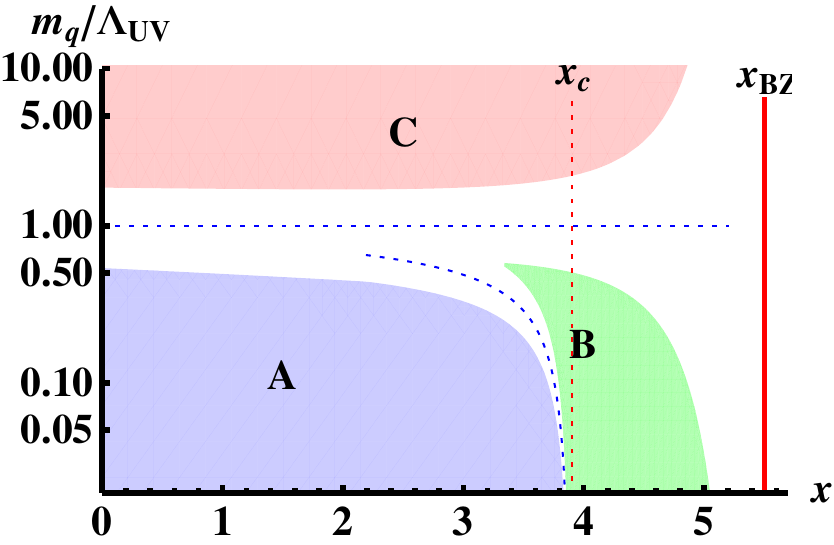}\hspace{10mm}%
\includegraphics[width=0.45\textwidth]{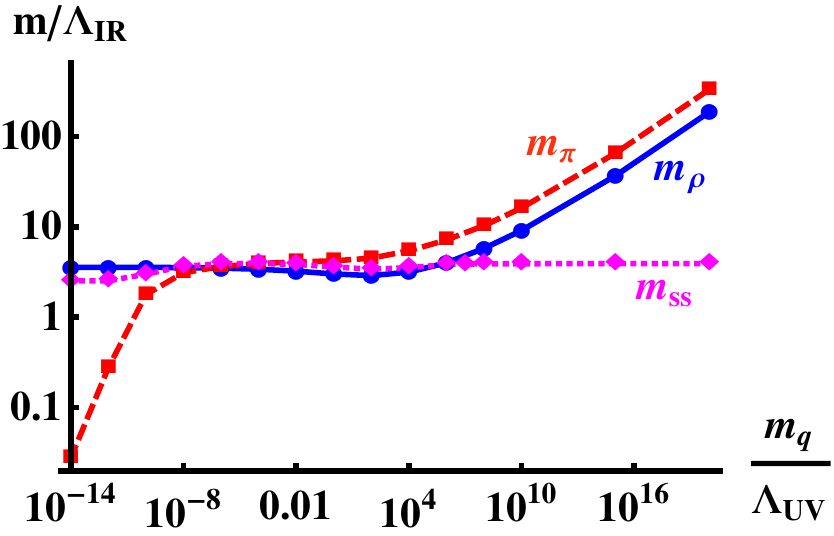}
\caption{Left: The regimes A, B, and C, where observables have qualitatively different dependence on the quark mass. The white regimes represent crossovers between the various scaling regimes. Right: The masses of the lowest vector (rho meson), pseudoscalar (pion), and singlet scalar states as a function of $m_q$ and in units of $\L_\mathrm{IR}$.}
\label{fig:scalingregimes}
\end{figure}

\subsection{Physics at finite quark mass and at finite $x$}

So far we concentrated on the phase diagram of V-QCD at zero quark mass, but there is also interesting behavior at finite quark mass $m_q$. When the quark mass is turned on, chiral symmetry is always broken, and the conformal transition changes to a crossover. Therefore there are no phase transitions on the $(x,m_q)$-plane for $0<x<\xBZ$. Instead, one can identify different regimes where observables (such as the chiral condensate and meson masses) have qualitatively different mass dependencies, see Fig.~\ref{fig:scalingregimes} (left)~\cite{Jarvinen:2015ofa}. In this diagram, the white regions represent crossovers between the different regimes, which are
\begin{itemize}
 \item[A] The regime where quark mass acts as a small (linear) perturbation to the theory at $m_q=0$.
 \item[B] The regime where the amount of walking in the model is controlled by the quark mass, leading to the so called ``hyperscaling''~\cite{DelDebbio:2010ze}.
 \item[C] The regime of large quark mass.
\end{itemize}
The results in regimes~A and~B are not sensitive to the details of the V-QCD action, but in regime~C there is quite a lot of dependence on the model. 
Interestingly, requiring match with QCD in regime~C is consistent with~\eqref{akasympt} which was obtained from considerations at zero $m_q$. In particular, both requiring linear meson trajectories at large excitation numbers, and a mass gap $\propto m_q$ at large $m_q$, independently imply that $a(\l)$ is constant.

The structure of Fig.~\ref{fig:scalingregimes} (left) can be verified, e.g., by studying the meson masses which are found by identifying the normalizable fluctuation modes of V-QCD around the vacuum solution. Full analysis requires including flavor structure and gauge fields in the TDBI action. There are both flavor singlet (corresponding to quark bilinear operators with the unit matrix $\mathbf{1}_{N_f\times N_f}$ in flavor space) and nonsinglet (corresponding to operators with the traceless Hermitean generators $t^a$) mesons.

In regime~B, the mass gap of the mesons and the condensate obey the standard hyperscaling relations
\be \label{hypersc}
 m_\mathrm{gap} \sim m_q^\frac{1}{\Delta_*} = m_q^\frac{1}{1+\gamma_*}\ , \quad \langle\bar qq\rangle \sim m_q^\frac{4-\Delta_*}{\Delta_*}=m_q^\frac{3-\gamma_*}{1+\gamma_*}
\ee
as $m_q \to 0$, which have also been studied in the dynamic AdS/QCD models~\cite{Evans:2014nfa}. Notice that regime~B also extends to $x<x_c$, where one should set $\gamma_*=\Delta_*-1=1$ in these formulae.

As an example of the dependence of the meson masses on $m_q$, we plot the pion (lowest nonsinglet pseudoscalar) mass $m_\pi$, the rho (lowest nonsinglet vector) mass $m_\rho$, and the mass of the lowest singlet scalar state $m_\mathrm{ss}$ as a function of $m_q$ in Fig.~\ref{fig:scalingregimes} (right). We fixed the value of $x$ right below the critical value $x_c$ such that the plot intersects all three regimes. In regime~A (lowest $m_q$) the pion mass is suppressed as required by the Gell-Mann-Oakes-Renner relation. In regime~B (intermediate $m_q$), all masses have the same dependence on $m_q$. The masses obey~\eqref{hypersc} in units of $\L_\mathrm{UV}$, but we have chosen to plot the masses in units of $\L_\mathrm{IR}$ so that the hyperscaling behavior is divided out. In regime~C (largest $m_q$) the scalar singlet state is a glueball and therefore its mass is suppressed with respect to the $\bar qq$ states by a factor of $m_q$.

\subsection{The S-parameter}

The S-parameter can be defined in terms of the difference of the vector-vector and axial-axial correlators. For a strongly interacting theory to be a viable candidate for the realization of technicolor, i.e., in order for the theory to pass the precision tests at LEP, this dimensionless parameter must be much smaller than one. It has been predicted that $S$ is suppressed in the walking regime~\cite{Appelquist:1998xf,Harada:2005ru,Kurachi:2006mu}, suggesting that walking technicolor theories are viable.

\begin{figure}[!tb]
\centering
\includegraphics[width=0.45\textwidth]{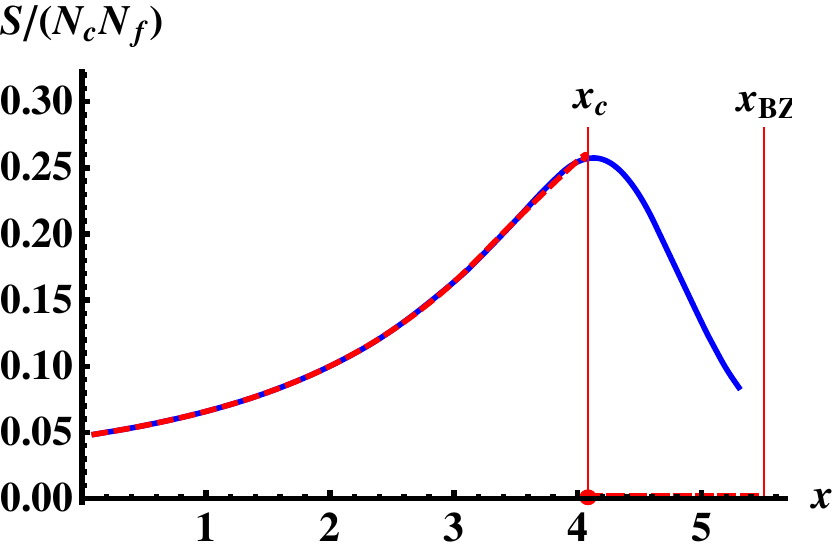}\hspace{10mm}%
\includegraphics[width=0.45\textwidth]{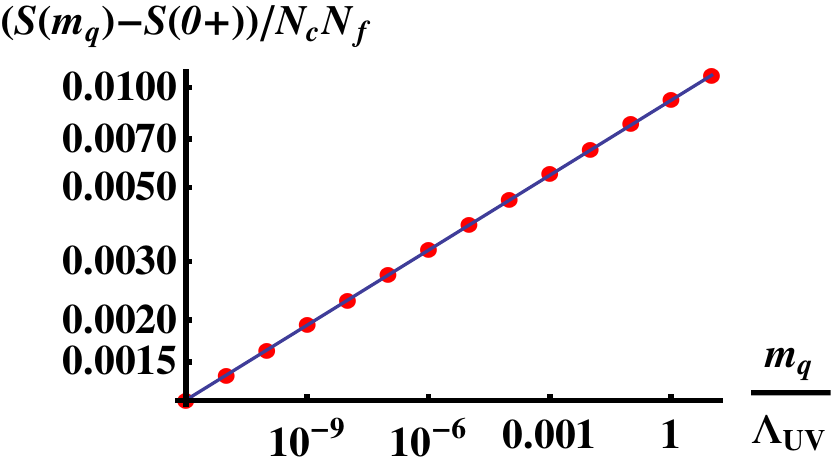}
\caption{Left: The S-parameter as a function of $x$ in V-QCD. The dashed red curve shows the value at zero quark mass, whereas the solid blue curve has $m_q=10^{-6}$. Right: The mass dependence of the S-parameter for $x=4.5$ in log-log scale. The red dots are the data and the blue line is a power-law fit.}
\label{fig:massesSparam}
\end{figure}

The (normalized) S-parameter can be extracted by analyzing the fluctuations of the V-QCD action and is shown as a function of $x$ in Fig.~\ref{fig:massesSparam} (left). The blue solid (red dashed) curves have $m_q=10^{-6}$ ($m_q=0$), respectively. The normalized S-parameter is enhanced rather than suppressed in the walking region (as $x \to x_c$ from below). For finite quark mass, the S-parameter is also nontrivial inside the conformal window~\cite{Jarvinen:2015ofa}, whereas at zero quark mass it vanishes there because chiral symmetry is intact. Notice that as a tiny quark mass is turned on in the conformal window, the S-parameter immediately jumps to a finite \order{N_fN_c} value. Such a discontinuity is consistent with recent field theory analysis~\cite{Sannino:2010ca}.

At small but finite values of $m_q$ inside the conformal window, the dependence of $S$ on $m_q$ is a power law (plus a constant), see Fig.~\ref{fig:massesSparam} (right)~\cite{Jarvinen:2015ofa}. The power turns out to be connected to the dimension $\Delta_{FF}$ of the ${\mathbb Tr} F^2$ operator at the fixed point:
\be \label{Sfit}
 S(m_q)-S(0^+) \sim m_q^{\frac{\Delta_{FF}-4}{\Delta_*}} \ ,\qquad (m_q \to 0)\ .
\ee

\begin{figure}[!tb]
\centering
\includegraphics[width=0.5\textwidth]{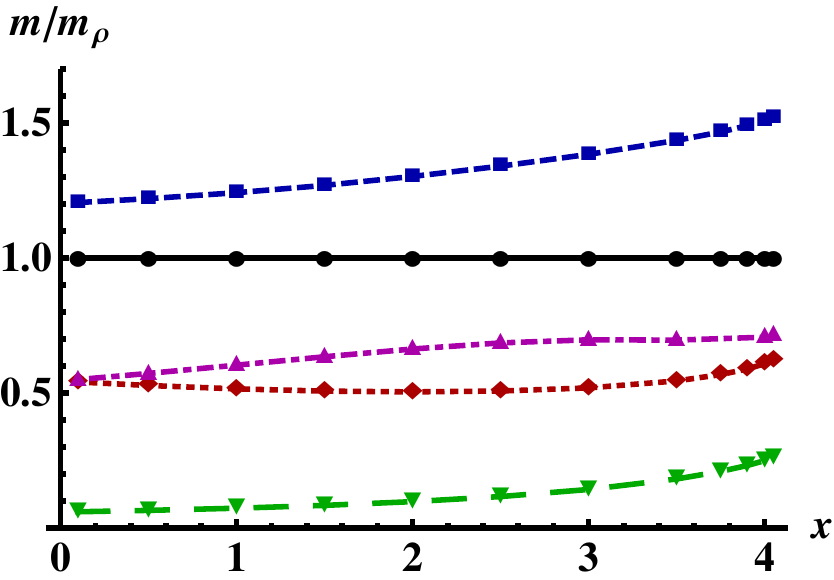}
\caption{The mass gaps of the different meson sectors normalized to the rho mass as a function of $x$. The black line is the rho mass, i.e., the vector mass gap. The blue dashed, red dotted, and magenta dot-dashed, curves are the axial, nonsinglet scalar, and singlet scalar mass gaps, respectively, while the green long-dashed curve is the pion decay constant.}
\label{fig:dilaton}
\end{figure}

\subsection{(Non)existence of a technidilaton mode}

It has been conjectured that the scalar singlet sector contains a ``dilaton'' as $x \to x_c$, an anomalously light Goldstone mode due to the breaking of conformal symmetry~\cite{Yamawaki:1985zg}. We plot the masses of the lowest states for the various meson sectors in V-QCD, normalized to the rho mass and as a function of $x$, in Fig.~\ref{fig:dilaton}. The scalar singlet mass gap is the magenta dot-dashed curve, which approaches a constant for $x \to x_c$, so there is no anomalously light technidilaton in V-QCD.

The example of V-QCD therefore shows that the mechanism with BF bound violation does not automatically yield a light technidilaton state (while it does yield Miransky scaling and hyperscaling relations). Interestingly, a detailed analysis~\cite{Alanen:2011hh,Arean:2013tja} shows that the scalar singlet sector is special. The fluctuation equation for the scalar singlets is ``critical'' for values of $r$ where the coupling walks, which suggests the presence of a light state. In the end it turns out, however, that the mass of the light state depends on the definition of the model in the deep IR, so that there is also room for model dependence. In V-QCD a relatively light, but not parametrically light, scalar is found as $x \to x_c$. In dynamic AdS/QCD models with a BKT transition, one does find a parametrically light scalar mode~\cite{Alho:2013dka}, but in the top-down $D3-D7$ defect model~\cite{Kutasov:2011fr} the results are qualitatively similar to V-QCD: there is a light, but no parametrically light scalar state.

\begin{figure}[!tb]
\centering
\includegraphics[width=0.45\textwidth]{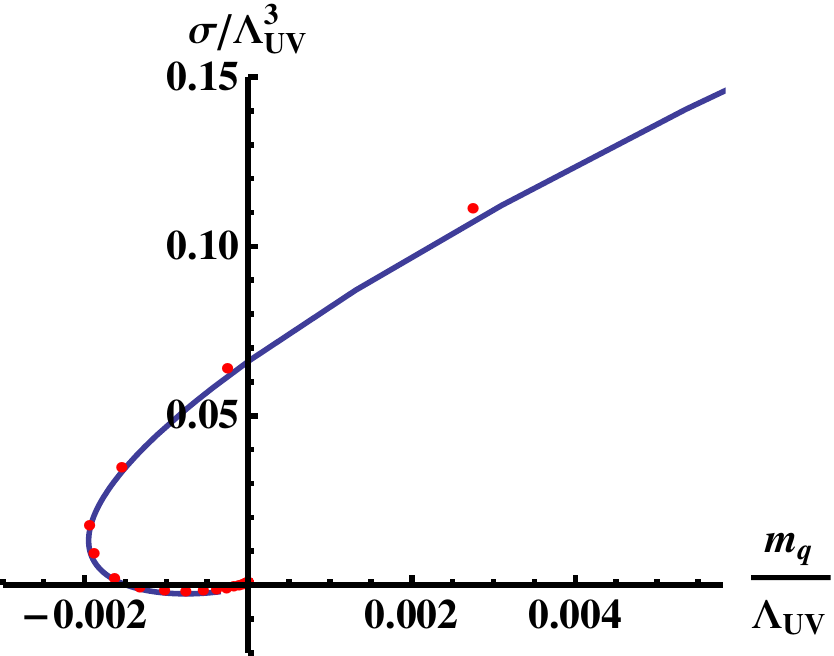}\hspace{10mm}%
\includegraphics[width=0.45\textwidth]{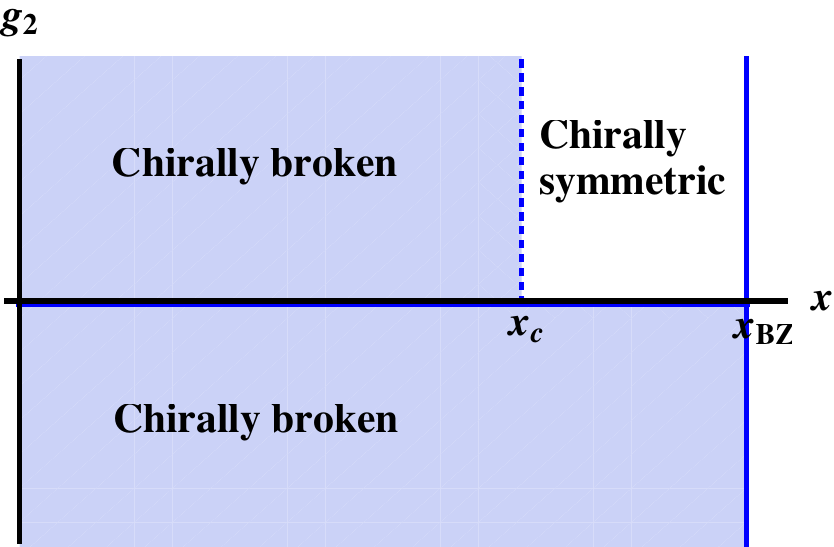}
\caption{Left: All regular V-QCD solutions on the $(m_q,\sigma)$-plane in the QCD regime. The red dots are data from V-QCD and the blue curve is an analytic fit.  Right: The phase diagram of V-QCD on the $(x,g_2)$-plane, where $g_2$ is the four-fermion coupling, at zero quark mass. }
\label{fig:efimov}
\end{figure}

\subsection{Efimov spirals and four fermion operators}

Finally we comment briefly on the dependence of the chiral condensate on the quark mass and how this is used to include four-fermion operators in the model. In the QCD regime, plotting the values of $m_q$ and the chiral condensate $\propto \sigma$ on the plane reveal a spiral~\cite{Iqbal:2011aj} structure, see Fig.~\ref{fig:efimov} (left)~\cite{Jarvinen:2015ofa}. (The spiral winds around the origin infinitely many times but this is not visible due to limited resolution.) We call this structure the Efimov spiral, and the additional solutions at fixed quark mass (which are found at intersection points of the spiral with a vertical line placed at the value of the quark mass) Efimov vacua. Such vacua are subdominant~\cite{Jarvinen:2011qe,Jarvinen:2015ofa}: the solution furthest away from the origin is the ``standard'' stable solution. In the conformal window there is no spiral, but $\sigma(m_q)$ is a simple curve.

The curve $\sigma(m_q)$ can be used to add four-fermion operators in the model by modifying the UV boundary conditions for the tachyon as suggested by Witten~\cite{Witten:2001ua}. The operator we want to add is
\be
 \propto g_2 \int d^4x \ (\bar qq)^2 \ .
\ee
The recipe is, that we first replace the coefficient $m_q$ by $\alpha$ in the UV expansion~\eqref{tachUV} for the tachyon, and then set the new boundary condition
\be
 \alpha = m_q + g_2 \sigma \ .
\ee
This means that all solutions at fixed $m_q$ are found by overlapping a tilted line instead of a vertical line with the Efimov spiral of Fig.~\ref{fig:efimov} and searching for the intersection points. Computing the free energy of each solution, we find for $m_q=0$ the phase diagram of Fig.~\ref{fig:efimov} (right). For $g_2>0$, there is no qualitative change with respect to the solutions at $g_2=0$, and there is still a BKT transition at $x=x_c$. But turning on any negative $g_2$ leads to an instability, a new vacuum appears which is energetically favorable with respect to the standard vacuum. Therefore there is a chirally broken phase for $g_2<0$ which is disconnected from the dominant vacuum at $g_2=0$. 

\section{Conclusions and outlook}

In this talk we have discussed results and predictions from the holographic V-QCD models, mostly near the conformal transition and in the conformal window. Many of these results appear universal, and may apply even slightly more in general than only in V-QCD. The outcome of the studies is encouraging: the model is consistent, at qualitative level, well with expectations based on field theory analyses.

Several additional extensions and applications of V-QCD are also under study. These include a study of the CP-odd sector (physics of the theta angle and axial anomaly of QCD)
 and finally fitting the model to QCD data, for example Yang-Mills thermodynamics from the lattice, experimental values of the QCD meson masses, and the equation of state at finite $x$ from the lattice. 

\section*{Acknowledgments}

The author wishes to thank K.~Yamawaki for the invitation to SCGT15 and the participants of the conference for several interesting discussions. This work was partially supported by the
``ARISTEIA II'' Action of the  ``Operational Programme Education and Lifelong Learning'' and was co-funded by the European Social Fund (ESF) and National Resources.

\bibliographystyle{ws-procs975x65}
\bibliography{scgt15-jarvinen}

\end{document}